\documentclass[sigconf]{acmart}
\AtBeginDocument{%
  }

\copyrightyear{2026}
\acmYear{2026}
\setcopyright{cc}
\setcctype{by}
\acmConference[CHI EA '26]{Extended Abstracts of the 2026 CHI Conference on Human Factors in Computing Systems}{April 13--17, 2026}{Barcelona, Spain}
\acmBooktitle{Extended Abstracts of the 2026 CHI Conference on Human Factors in Computing Systems (CHI EA '26), April 13--17, 2026, Barcelona, Spain}
\acmDOI{10.1145/3772363.3798570}
\acmISBN{979-8-4007-2281-3/2026/04}

\usepackage{enumitem}
\usepackage{graphicx}

\usepackage{booktabs}
\usepackage{siunitx}
\newcommand{\PH}[1]{\textbf{#1}}
\newcommand{\KappaPH}{\PH{0.78}}
\newcommand{\KappaNPH}{\PH{96}}
\newcommand{\VendFOneLoPH}{\PH{0.80}}
\newcommand{\VendFOneHiPH}{\PH{0.86}}
\newcommand{\BalDiffPH}{\PH{0.05}}
\newcommand{\JSerrTopPH}{\PH{27\%}}

\begin{document}

\title{ConsentDiff at Scale: Longitudinal Audits of Web Privacy Policy Changes and UI Frictions}

\author{Haoze Guo}
\email{hguo246@wisc.edu}
\orcid{0009-0009-5987-1832}
\affiliation{%
  \institution{University of Wisconsin - Madison}
  \city{Madison}
  \state{WI}
  \country{USA}
}

\renewcommand{\shortauthors}{Haoze Guo}

\begin{abstract}
Web privacy is experienced via two public artifacts: site utterances in policy texts, and the actions users are required to take during consent interfaces. In the extensive cross-section audits we've studied, there is a lack of longitudinal data detailing how these artifacts are changing together, and if interfaces are actually doing what they promise in policy. \textbf{ConsentDiff} provides that longitudinal view. We build a reproducible pipeline that snapshots sites every month, semantically aligns policy clauses to track clause-level churn, and classifies consent-UI patterns by pulling together DOM signals with cues provided by screenshots. We operationalize claim–UI alignment by mapping policy claims to observable UI predicates, connecting common policy claims to observable predicates, and enabling comparisons over time, regions, and verticals. Our measurements suggest continued policy churn, systematic changes to eliminate a higher-friction banner design, and significantly higher alignment where rejecting is visible and lower friction.
\end{abstract}



\maketitle

\section{Introduction}
Online privacy is enacted through two public-facing artifacts: (i) \emph{privacy policies}, which disclose purposes, retention, and sharing, and (ii) \emph{consent interfaces} (CMP banners/modals), which implement user choice. Prior research has documented pervasive violations and dark-pattern designs in cookie banners and consent flows \cite{nouwens2020cookiebanners,matte2020cookierespect,mathur2019darkpatterns}. However, we still lack a longitudinal, web-scale view that jointly tracks \emph{what sites say} in policy text and \emph{what users must do} in consent interfaces, along with a concrete metric for whether these two artifacts align.

This is a problem for both end users and regulators. Policies are often updated without any notice, whether due to template refreshes, vendor switches, or jurisdiction-specific edits. Consent banners also change over time through A/B testing, CMP defaults, and implementation updates. Point-in-time audits can identify obvious violations, but they often miss \emph{how} deployments evolve over time, which patterns persist, and whether policy claims keep pace with UI changes. At the same time, regulatory and industry guidance (e.g., recommendations for valid consent flows \cite{edpb2020consent,iab2020tcf}) describe what compliant consent \emph{should} look like, but they do not provide a reproducible way to measure whether common policy claims (e.g., ``opt-in,'' ``easy reject'') correspond to the actual actions users must take.

We present \textbf{ConsentDiff}, a measurement framework that jointly tracks policy text and consent interfaces over time. Each month, we snapshot a stratified set of sites, semantically align policy clauses across versions, classify consent-UI patterns from DOM and screenshots, and compute a \emph{claim--UI alignment} score that pairs policy statements with observable defaults and actions. Our work complements research on temporal drift in privacy recall which models how people misremember original sharing audience \cite{guo2026temporaldriftprivacyrecall}; our focus is on the \emph{site-side} artefacts, as well as noting their evolution.

We study how policy text and consent interfaces co-evolve at scale, guided by three research questions:
\begin{enumerate}[leftmargin=*,noitemsep,topsep=0pt]
  \item[\textbf{RQ1:}] \textbf{Claim--action consistency.} When privacy policies make user-choice claims (e.g., easy rejection, opt-in, withdrawal), how often do the observed consent interfaces satisfy the corresponding UI predicates?
  \item[\textbf{RQ2:}] \textbf{Longitudinal drift.} How do policy clauses, consent-UI patterns, and claim--UI alignment change over time across regions and verticals?
  \item[\textbf{RQ3:}] \textbf{Robust interpretation.} How sensitive are longitudinal comparisons to imperfect banner surfacing and classifier error, and what diagnostics bound these risks?
\end{enumerate}

\section{Related Work}

\subsection{Consent Interfaces and Dark Patterns}
Large-scale audits have documented pervasive lack of compliance with and the use of dark-pattern designs in consent banners and CMP deployments, which illustrate how choices in interface can nudge users towards acceptance and obscure paths to rejection \cite{nouwens2020cookiebanners,matte2020cookierespect,mathur2019darkpatterns,degeling2019wevalue}. Practice and policy guidance add further specificity to valid consent elements (e.g., unambiguous opt-in, symmetry of choices, easy withdrawal) and have codified vendor-facing frameworks that many sites adopt \cite{edpb2020consent,iab2020tcf}. Outside of consent dialogs, dark-pattern taxonomies and crawls have captured patterns of manipulative designs across e-commerce and platforms, including intrusion, default bias, and interference with interfaces \cite{mathur2019darkpatterns,gray2021}. Our focus on claim--UI alignment also relates to emerging HCI work on user-facing transparency cues in algorithmic systems, which similarly treats visible interface elements as measurable signals of what platforms reveal (or obscure) about system behavior and user control \cite{guo2026feedtaxonomyuserfacingcues}. In our setting, we apply a similar interface-first lens to consent and privacy interfaces, but center whether user-choice claims in policy text are matched by observable, low-friction actions in the UI.

\subsection{Automated Policy Analysis and Web-Scale Measurement}
Automated tools segment and label privacy policies to enable designed query and compliance studies on a massive scale \cite{harkous2018polisis,lippi2019claudette}. Extending this area of work, we align policy \emph{clauses over time} to measure clause-level churn, and to relate this churn to consent UI changes. There are research-oriented rankings that utilize methodology to reduce "bias" in the top lists for our sampling and instrumentation \cite{pochat2019tranco}, which is supported with classic web-tracking studies on the importance of maintaining long term request/cookie logs and distinguishing between first/third parties, capturing screenshots indicating the UI states during data collection methods \cite{englehardt2016online,acar2014webneverforgets,lerner2016raiders}. More broadly, our work sits in a line of web-facing measurement that treats public web artifacts as inputs to automated analysis pipelines and stresses the importance of robustness to real-world web variability and failure modes \cite{guo2026hiddeninplaintextbenchmarksocialwebindirect}.

\section{Method}
\subsection{Data and Sampling}
We produced a stratified frame seeded by Tranco, balanced by rank and vertical (news, retail, social, video) and bucketing regions as EU, US-CA, and Other for jurisdictional comparisons \cite{pochat2019tranco}. A headless browser, with new profile for each site, loads the homepage and privacy-policy page with geo/language hints to surface banners. For each snapshot, we retain: raw HTML, serialized consent-DOM subtree, full page screenshots of banners/modals, steps if they occur, and compact network log capturing request/cookie events with first/third-party flags. We clear storage between runs; the artefacts are content-hashed and time-stamped for de-duplication and reproducibility.

\emph{Banner Elicitation Settings.}
Each visit used a fresh Chrome profile (viewport 1366$\times$768), regional \texttt{Accept-Language}, and region-specific IP. Storage was cleared, service-worker reuse blocked, and we waited up to 10\,s for late-loading banners. Banner-surfacing failure (no banner despite region bucket) was 19.0\% (EU), 48.0\% (US--CA), and 43.0\% (Other) over successful loads; these flags are included in the released aggregates.

\subsection{Policy and UI Analysis}
\emph{Policy clauses.} Each policy is parsed into clause candidates utilizing headings and sentence boundaries. Each clause, is sidled with similar embedded clauses across subsequent snapshots with a cost that incorporates semantic and edit similarity \cite{reimers2019sbert,levenshtein1966}. The clauses are labeled with a compact taxonomy (Purpose, Retention, Sharing, LegalBasis, Rights, Contact), and we consider the amount of \emph{clause churn} as the share of clauses that substantially differ between months.

\emph{Policy clause extraction/alignment evaluation.} In addition to the UI-pattern gold set, we evaluated policy processing on a hand-labeled subset of policy snapshots. We assessed (i) clause boundary quality against manual segmentation and (ii) month-to-month clause alignment accuracy on manually matched clause pairs spanning minor rewrites and template edits. Clause extraction was most reliable for heading-structured policies, while alignment errors concentrated in large template refreshes and pages with repeated boilerplate. We use these diagnostics to interpret clause-churn estimates conservatively and emphasize taxonomy-level trends over individual clause trajectories.

\emph{Consent patterns.} Consent UIs are classified by merging documented features (e.g., scrollable containers, toggle default selections, step indicators, calling text primary/secondary buttons/options, flags as visible or not) with screenshot markers. Weak rules confer training labels while a lighter classifier, a pooling image representation of documented features concatenated, predicts one of \textsc{Scroll-Wall}, \textsc{Accordion}, \textsc{Multi-Step}, \textsc{Pre-ticked}, or \textsc{Reject-Hidden} \cite{ratner2017snorkel}. A small, hand-labeled set aids in calibration and reporting precision/recall. We then calculate longitudinal prevalence and the row-normalized transition matrix (Table~\ref{tab:transitions}).

\emph{Gold set and agreement.}
We evaluate the UI-pattern classifier on a gold set of \textbf{240} snapshots, stratified by region and vertical (news, retail, social, video) with \textbf{20} samples per cell (12 cells total). A random subset of \KappaNPH{} items was double-coded using a shared codebook; inter-annotator agreement was Cohen's $\kappa=\KappaPH$, with disagreements resolved by adjudication.

\subsection{Metrics and Evaluation}
\emph{Claim--UI alignment.} We pair policy claims with necessary UI predicates (e.g., default-off, visible ``Reject all'', steps-to-reject $\le 2$) to obtain an alignment score $A\in[0,1]$ per snapshot. We summarize $A$ by region and vertical (Figure~\ref{fig:alignment_dist}) and relate it to observed pattern shares.
\emph{Longitudinal analysis.} In the monthly series, we track the churn as well as the prevalence of patterns; we flag structural shifts with simple change point checks \cite{killick2012changepoint}. We summarize event responses as difference-in-differences with site and month fixed effects and use clustered inference \cite{angrist2009mhe,wooldridge2010panel}. Quality controls involve short-interval re-crawls for stability, and we report classifier metrics on the gold set.

\subsection{Robustness Checks}
We evaluate robustness along four axes: policy-processing reliability, UI-pattern classification error, banner non-surfacing, and sampling sensitivity. Manual checks on clause segmentation and temporal alignment indicate that policy-processing errors are concentrated in major template refreshes and repeated boilerplate; accordingly, we interpret clause churn conservatively and emphasize taxonomy-level trends.

For consent-UI classification, the gold-set evaluation yields macro-F1$\approx$0.84 (Table~\ref{tab:clf}), with stress checks for off-viewport reject buttons, image-heavy banners, and localization. Stratifying by major CMP vendor families yields similar macro-F1 (\VendFOneLoPH{}--\VendFOneHiPH{}), suggesting limited vendor-specific overfitting. Propagating the observed confusion matrix via 1{,}000 bootstrap draws perturbs median $A$ by $\le 0.02$ and DiD coefficients by $\le 0.01$, leaving signs and group orderings unchanged.

Banner surfacing is asymmetric by region, so we report inverse-probability-weighted (IPW) estimates and a worst-case sensitivity bound. Banner non-surfacing is associated with script failures and late-loading UI; \JSerrTopPH{} of non-surfacing snapshots show JS console errors during elicitation, and IPW achieves covariate balance with max standardized mean difference \BalDiffPH{}. Under a worst-case MNAR bound, the EU--US gap shrinks by at most 0.02, with signs and relative orderings preserved.

Finally, Tranco-stratified estimates remain stable under rank weighting and inverse-probability weighting for banner surfacing by region, with effect directions and relative magnitudes unchanged.

\section{Measurement and Findings}

\subsection{Scale and Coverage}
Our frame contains \textbf{2{,}400} domains tracked over \textbf{9} monthly waves (\textbf{21{,}600} site–month snapshots). Region buckets: EU (\textbf{900} domains; 8{,}100 snaps), US--CA (\textbf{1{,}000}; 9{,}000), Other (\textbf{500}; 4{,}500). We detect a consent banner in \textbf{13{,}248/21{,}600} snapshots (\textbf{61.4\%}); by region: EU \textbf{6{,}156/8{,}100} (\textbf{76.1\%}), US--CA \textbf{4{,}230/9{,}000} (\textbf{47.0\%}), Other \textbf{2{,}862/4{,}500} (\textbf{63.6\%}). Policy text fetch coverage is \textbf{19{,}958/21{,}600} (\textbf{92.4\%}), screenshots \textbf{20{,}693/21{,}600} (\textbf{95.8\%}), and serialized consent-DOM subtrees \textbf{20{,}105/21{,}600} (\textbf{93.1\%}); analyses use per-artifact denominators.

\subsection{Effect Sizes and Robustness}
\emph{Group gaps.} Median $A$ is higher in EU than US--CA by 0.09 (95\% CI [0.07, 0.11]; Cliff's $\delta=0.34$); News/Social exceed Retail. Visible “Reject all” associates with +0.12 $A$ [0.11, 0.13]; steps-to-reject $\le 2$ adds +0.07 [0.05, 0.08]; default-off +0.05 [0.04, 0.06].
\emph{Event responses.} Post-enforcement we observe: “Reject all” +9.3 pp [6.8, 11.7], \textsc{Pre-ticked} -4.8 pp [-6.2, -3.4], and $A$ +0.04 [0.03, 0.05]; pre-trend tests are null (joint $p=0.41$).
\emph{Robustness.} Findings persist under rank weighting, excluding top–50 sites, alternate windows (±2/±3 months), EEA+UK bucketing, and vendor fixed effects (magnitudes within 0.01–0.02 of main).

\subsection{Trends, Alignment, and Event Effects}
We track monthly clause churn by taxonomy alongside consent-UI pattern prevalence; Figure~\ref{fig:alignment_dist} shows cross-sectional $A$ by region/vertical. We observe: sustained churn in \textsc{Purpose}/\textsc{Sharing} in Q3–Q4, a gradual shift from \textsc{Scroll-Wall} toward \textsc{Accordion}, and a decline in \textsc{Pre-ticked} in EU buckets—consistent with template iteration and incremental CMP-layer adjustments.

Table~\ref{tab:transitions} reports row-normalized transitions between consent-UI patterns: \textsc{Scroll-Wall} is most persistent (with spillover to \textsc{Accordion}); \textsc{Reject-Hidden} is comparatively unstable, often moving to \textsc{Multi-Step}/\textsc{Accordion}. Following dated enforcement announcements, we see level shifts: visible “Reject all” rises, \textsc{Pre-ticked} falls, and alignment $A$ increases; placebo pre-trends are null. Overall, higher $A$ is associated with visible “Reject all,” default-off toggles, and $\leq\!2$ steps-to-reject.

\begin{table}[t]
  \centering
  \small
  \setlength{\tabcolsep}{3pt}
  \caption{Per-class UI-pattern classifier performance on the 240-sample test set.}
  \label{tab:clf}
  \begin{tabular}{lccc}
    \toprule
    \textbf{Class} & \textbf{Precision} & \textbf{Recall} & \textbf{F1} \\
    \midrule
    \textsc{Scroll-Wall}    & 0.91 (0.88--0.94) & 0.89 (0.85--0.92) & 0.90 (0.87--0.93) \\
    \textsc{Accordion}      & 0.88 (0.84--0.91) & 0.90 (0.86--0.93) & 0.89 (0.86--0.92) \\
    \textsc{Multi-Step}     & 0.85 (0.80--0.90) & 0.83 (0.77--0.88) & 0.84 (0.79--0.88) \\
    \textsc{Pre-ticked}     & 0.82 (0.76--0.88) & 0.78 (0.71--0.84) & 0.80 (0.73--0.85) \\
    \textsc{Reject-Hidden}  & 0.80 (0.73--0.86) & 0.76 (0.69--0.83) & 0.78 (0.71--0.83) \\
    \midrule
    \textbf{Macro Avg.}     & \textbf{0.85} & \textbf{0.83} & \textbf{0.84} \\
    \bottomrule
  \end{tabular}
\end{table}

\subsection{Pattern Dynamics and Alignment}
\begin{figure}[t]
  \centering
  \includegraphics[width=\linewidth]{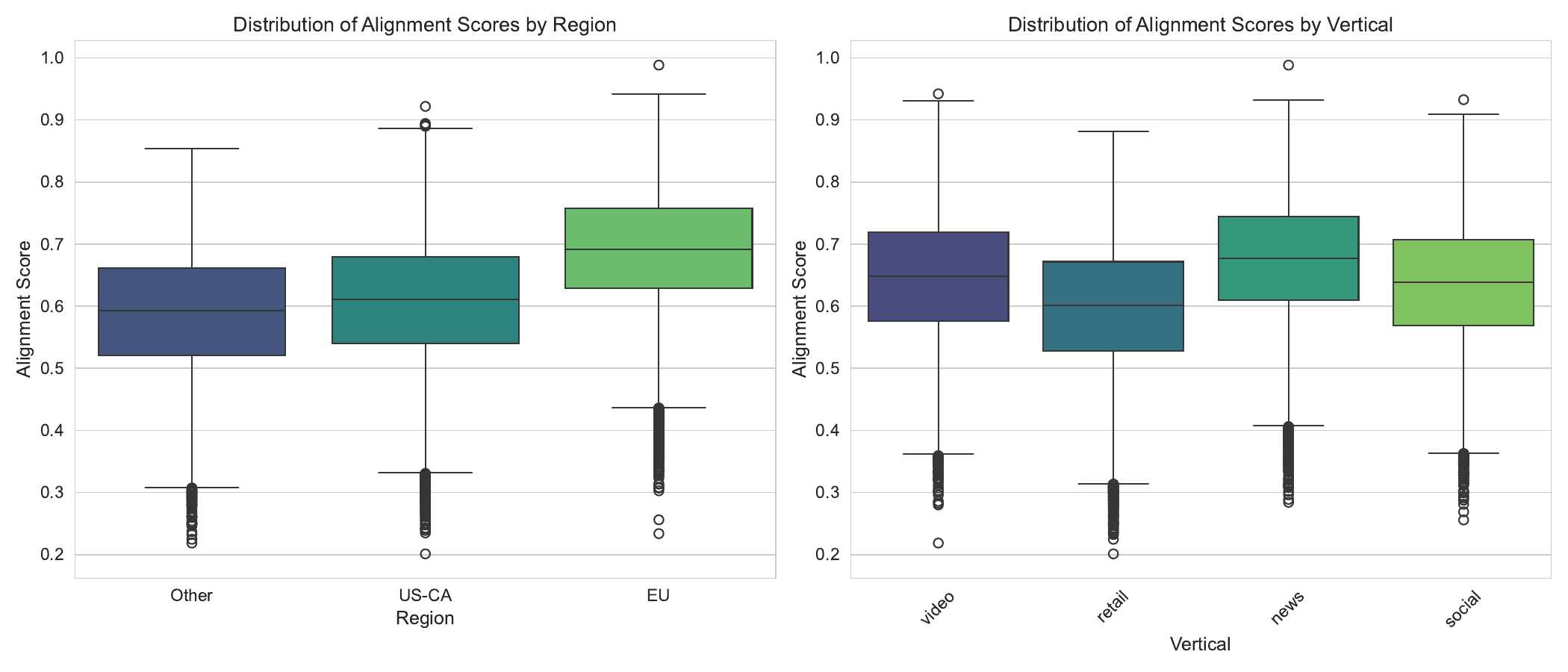}
  \caption{Alignment score distributions by region (top) and vertical (bottom). Boxes show medians and IQR; whiskers extend to 1.5$\times$IQR.}
  \label{fig:alignment_dist}
\end{figure}

Table~\ref{tab:transitions} reports row-normalized month-to-month transition probabilities between predicted consent-UI patterns. \textsc{Scroll-Wall} exhibits the highest persistence with notable transitions to \textsc{Accordion}; \textsc{Reject-Hidden} is comparatively unstable and frequently moves toward \textsc{Multi-Step} or \textsc{Accordion}. Figure~\ref{fig:alignment_dist} summarizes the claim--UI alignment score by region and vertical: EU-bucketed sites show higher medians than US-CA/Other, driven by visible “Reject all” and fewer steps-to-reject; retail underperforms news/social due to more frequent default-on toggles.

\paragraph{Alignment score $A$.}
For snapshot $i$, we extract binary \emph{claims} $\mathcal{C}_i$ from policy text and binary \emph{UI predicates} $\mathcal{U}_i$ from the banner/flow. We evaluate the implication set
$\mathcal{M} = \{(c,u,w)\}$, where $c\in\mathcal{C}_i$ (e.g., \textit{opt-in}, \textit{minimization}, \textit{easy reject}), $u\in\mathcal{U}_i$ (e.g., default-off, visible ``Reject all'', steps-to-reject $\le 2$), and $w$ is a nonnegative weight s.t.\ $\sum_{(c,u,w)\in\mathcal{M}} w=1$. We define
\[
A_i \;=\; \sum_{(c,u,w)\in \mathcal{M}} w \cdot \mathbf{1}[c \Rightarrow u],
\]
We treat undetected predicates as unsatisfied, yielding a conservative lower-bound estimate of alignment under imperfect observability. We use $w$ = \{Reject-all visibility: \textbf{0.4}; Default-off: \textbf{0.3}; Steps-to-reject $\le 2$: \textbf{0.2}; Reopen/withdrawal affordance: \textbf{0.1}\}.

\begin{table}[t]
  \centering
  \small
  \setlength{\tabcolsep}{6pt}
  \caption{Ablation on $A$: median change when dropping one predicate at a time (pp = percentage points).}
  \label{tab:ablation}
  \begin{tabular}{lcc}
    \toprule
    \textbf{Predicate removed} & \textbf{$\Delta$ median $A$ (pp)} & \textbf{Cliff's $\delta$} \\
    \midrule
    Visible ``Reject all''       & $-6.1$ [\,$-7.3,-5.0$\,] & 0.29 \\
    Default-off toggles          & $-5.0$ [\,$-6.0,-4.0$\,] & 0.23 \\
    Steps-to-reject $\le 2$      & $-3.2$ [\,$-4.1,-2.4$\,] & 0.18 \\
    Reopen/withdrawal affordance & $-1.7$ [\,$-2.3,-1.2$\,] & 0.09 \\
    \bottomrule
  \end{tabular}
\end{table}

To quantify these gaps, we summarize alignment by group in Figure~\ref{fig:alignment_dist} and report predicate influence via the ablation in Table~\ref{tab:ablation}. Median alignment is higher where a visible “Reject all” is present and when steps-to-reject $\le 2$. We also estimate a two-way fixed-effects model at the site-month level
\[
A_{it} \;=\; \alpha \;+\; \beta_1 \mathrm{EU}_i \;+\; \beta_2 \mathrm{Retail}_i \;+\; \boldsymbol{\delta}^\top \mathbf{1}\{\text{pattern}_{it}\} \;+\; \mu_i \;+\; \tau_t \;+\; \varepsilon_{it},
\]
with site fixed effects $\mu_i$ and month fixed effects $\tau_t$. Coefficients on EU are positive and significant, while retail is negative; pattern indicators align with the descriptive ordering (\textsc{Accordion} $>$ \textsc{Scroll-Wall} $>$ \textsc{Multi-Step} $>$ \textsc{Reject-Hidden} $>$ \textsc{Pre-ticked}). Results are stable to winsorizing $A$ at the 1st/99th percentiles and to re-weighting by site rank.

\begin{table}[t]
  \centering
  \small
  \setlength{\tabcolsep}{5pt}
  \renewcommand{\arraystretch}{1.2}
  \caption{Row-normalized transitions of consent-UI patterns.}
  \label{tab:transitions}
  \begin{tabular}{l
                  S[table-format=2.1]
                  S[table-format=2.1]
                  S[table-format=2.1]
                  S[table-format=2.1]
                  S[table-format=2.1]}
    \toprule
    \textbf{Prev.\,$\rightarrow$\,Next}
      & {\textsc{SW}}
      & {\textsc{ACC}}
      & {\textsc{MS}}
      & {\textsc{PT}}
      & {\textsc{RH}} \\
    \midrule
    \textsc{Scroll-Wall}   & \bfseries 72.3 & 14.8 & 6.1 & 3.2 & 3.6 \\
    \textsc{Accordion}     & 18.5 & \bfseries 65.0 & 9.2 & 3.7 & 3.6 \\
    \textsc{Multi-Step}    & 10.1 & 21.4 & \bfseries 58.7 & 4.5 & 5.3 \\
    \textsc{Pre-ticked}    & 8.3  & 16.0 & 7.8  & \bfseries 60.9 & 7.0 \\
    \textsc{Reject-Hidden} & 9.0  & 12.7 & 10.5 & 6.9  & \bfseries 60.9 \\
    \bottomrule
  \end{tabular}

  \vspace{0.2em}
  \parbox{0.96\linewidth}{\footnotesize SW{=}Scroll-Wall, ACC{=}Accordion, MS{=}Multi-Step, PT{=}Pre-ticked, RH{=}Reject-Hidden.}
\end{table}

\section{Discussion}
For regulators, longitudinal artefact-level audits complement one-off enforcement by showing \emph{when} and \emph{how} sites change both claims and consent frictions. The claim--UI alignment score $A$ can triage investigations by surfacing low-$A$ domains or sudden drops around events. For CMP vendors and site operators, the transition patterns (Table~\ref{tab:transitions}) highlight how deployments evolve under product and policy constraints. Concrete remedies include presenting a visible ``Reject all'' co-equal with ``Accept all'', default-off for non-essential purposes, and keeping steps-to-reject $\leq 2$ to reduce friction and improve alignment.

Even with stratification based on Tranco, sampling might miss long-tail or niche verticals. In cases where the geo/language heuristics do not elicit some of the banners, you may see some extension of the locale variance and A/B testing noise. The DOM+vision classifier we describe in this paper may label some edge cases incorrectly; we report precision and recall on a gold set, and provide labels with which others may replicate our work. Policy parsing may drift due to template changes; our semantic alignment and stability checks mitigate that drift but do not eliminate it completely. Just as our network observations only reflect behavior in the context of our own sessions, we are unable to capture server-side differences.

\section{Conclusion}
\textit{ConsentDiff} presents an efficient, robust way to understand what's in privacy policies and consent interfaces at the same time and on a web-scale. It enables reproducible, longitudinal audits and pre/post comparisons across regions and verticals. It can also evaluate CMP/UI interventions by tracking shifts in predicted patterns and alignment. By combining clause-level policy diffs with a minimal DOM + vision classifier of consent interface patterns and claim-UI alignment score, we quantify temporal trends, transition dynamics, and jurisdictional differences without a user study. We find evidence for stable template implementations in policies; systematic differences in banners; and a higher degree of alignment where visible "Reject all" and low steps-to-reject interfaces exist. Future work will expand to other privacy-control surfaces beyond CMP banners, scale multimodal classifiers with larger stratified gold sets, and strengthen causal identification of event-aligned changes via richer event timing, alternative controls, and triangulation with independent evidence.

\bibliographystyle{ACM-Reference-Format}
\bibliography{refs}

\end{document}